\pdfoutput=1
\documentclass[preprint,superscriptaddress,11pt]{revtex4-1}

\usepackage{geometry,graphicx,color}
\usepackage{amsmath,mathtools,amssymb,mathrsfs}
\usepackage{dcolumn}
\usepackage{bm,extarrows,ulem,cancel}
\usepackage{slashed}
\usepackage[utf8]{inputenc}
\usepackage{CJKutf8}

\newcommand{\be}{\begin{equation}}
\newcommand{\ee}{\end{equation}}
\newcommand{\ba}{\begin{eqnarray}}
\newcommand{\ea}{\end{eqnarray}}
\newcommand{\bd}{\begin{cases}}
\newcommand{\ed}{\end{cases}}
\newcommand{\no}{\nonumber \\}
\newcommand{\gsim}{\mathrel{\hbox{\rlap{\lower.55ex \hbox {$\sim$}}
                   \kern-.3em \raise.4ex \hbox{$>$}}}}
\newcommand{\lsim}{\mathrel{\hbox{\rlap{\lower.55ex \hbox {$\sim$}}
                   \kern-.3em \raise.4ex \hbox{$<$}}}}

\def\roughly#1{\mathrel{\raise.3ex\hbox{$#1$\kern-.75em%
\lower1ex\hbox{$\sim$}}}}
\def\lsim{\roughly<}
\def\gsim{\roughly>}
\def\fm{{\mbox{fm}}}

\def\({\left(}
\def\){\right)}
\def\[{\left[}
\def\]{\right]}
\def\<{\langle}
\def\>{\rangle}
\def\lag{\langle}
\def\rag{\rangle}
\def\pd{\partial}

\def\1{\left}
\def\2{\right}

\def\l{{\lambda}}

\def\d{{\delta}}
\def\D{{\Delta}}

\def\e{{\epsilon}}

\def\a{{\alpha}}
\def\b{{\beta}}

\def\G{{\Gamma}}
\def\h{{\eta}}
\def\p{{\pi}}

\def\m{{\mu}}
\def\n{{\nu}}
\def\r{{\rho}}
\def\s{{\sigma}}

\def\t{{\tau}}

\def\x{{\xi}}

\def\et{{\eta}}
\def\fm{{\text{fm}}}
\def\cs{{\text{CS}}}
\def\mev{{\text{MeV}}}
\def\gev{{\text{GeV}}}

\newcommand{\dif}{\,\mathrm{d}}

\date{\today}

\begin{document}

\title{{\bf Chiral Magnetic Effect in Isobar Collisions from Stochastic Hydrodynamics}}
\thanks{The work of G.R.L is supported by International Program for Ph.D Candidates, Sun Yat-Sen University. The work of J.~L. is supported by the National Science Foundation under Grant No.PHY-1913729. The work of S.L. is supported by NSFC under Grant Nos 11675274 and 11735007. The work of L.Y is supported by Shanghai Pujiang Program No.19PJ1401400.}

\author{Gui-Rong Liang (\begin{CJK*}{UTF8}{gbsn}梁桂荣\end{CJK*})}
\email{Email address: bluelgr@sina.com}
\affiliation{School of Physics and Astronomy, Sun Yat-Sen University, Zhuhai, 519082, China}

\author{Jinfeng Liao (\begin{CJK*}{UTF8}{gbsn}廖劲峰\end{CJK*})}
\affiliation{Physics Department and Center for Exploration of Energy and Matter, Indiana University, 2401 N Milo B. Sampson Lane, Bloomington, Indiana 47408, USA}

\author{Shu Lin (\begin{CJK*}{UTF8}{gbsn}林树\end{CJK*})}
\affiliation{School of Physics and Astronomy, Sun Yat-Sen University, Zhuhai, 519082, China}

\author{Li Yan (\begin{CJK*}{UTF8}{gbsn}严力\end{CJK*})}
\affiliation{Key laboratory of Nuclear Physics and Ion-beam Application (MOE) \& Institute of Modern Physics, Fudan University, 220 Handan Road, 200433, Yangpu District}

\author{Miao Li (\begin{CJK*}{UTF8}{gbsn}李淼\end{CJK*})}
\affiliation{Department of Physics, Southern University of Science and Technology, Shenzhen 518055, China}

\begin{abstract}

We study chiral magnetic effect in collisions of AuAu, RuRu and ZrZr at $\sqrt{s}=200\gev$. The axial charge evolution is modeled with stochastic hydrodynamics and geometrical quantities are calculated with Monte Carlo Glauber model. By adjusting the relaxation time of magnetic field, we find our results in good agreement with background subtracted data for AuAu collisions at the same energy. We also make prediction for RuRu and ZrZr collisions. We find a weak centrality dependence of initial chiral imbalance, which implies the centrality dependence of chiral magnetic effect signal comes mainly from those of magnetic field and volume factor. Our results also show an unexpected dependence on system size: while the system of AuAu has larger chiral imbalance and magnetic field, it turns out to have smaller signal for chiral magnetic effect due to the larger volume suppression factor.\\


\end{abstract}

\maketitle

\newpage

\section{Introduction}

The anomalous transport of chiral magnetic effect (CME) has gained significant attention over the past few years \cite{Kharzeev:2007jp,Fukushima:2008xe}. If local parity odd domain is present in quark-gluon plasma produced in heavy ion collisions, CME leads to charge separation along the magnetic field generated in off-central collisions:
\begin{align}\label{cme}
{\bf j}_e=\sum_f\frac{N_cq_f^2}{2\pi^2}\m_A{\bf B},
\end{align}
with the chiral imbalance $\m_A$ characterizing local parity violation. This offers the possibility of detecting local parity violation in quantum chromodynamics (QCD). The charge separation has been actively searched experimentally \cite{Adamczyk:2013hsi,Abelev:2012pa,Adamczyk:2014mzf}. However, we are still far from consensus in the status of CME largely due to the difficulty in determine CME both experimentally and theoretically, see \cite{Kharzeev:2015znc,Zhao:2019hta,Huang:2015oca,Liao:2014ava} for recent reviews. Experimentally, charge separation needs to be measured through charged hadron correlation on event-by-event basis. Unfortunately the charged hadron correlation is dominated by flow related background with different possible origins \cite{Schlichting:2010qia,Bzdak:2010fd,Wang:2009kd}. Different observables and experimental techniques have been proposed and implemented to exclude flow related background \cite{Bzdak:2012ia,Wen:2016zic,Xu:2017qfs,Magdy:2017yje}. In addition, STAR collaboration proposes to search for CME in isobar collisions \cite{Skokov:2016yrj}. Since the isobar contain the same atomic number but different proton numbers, the corresponding collisions are supposed to generate the same flow background but different magnetic field and thus different charge separation, providing an unambiguous way of distinguishing the CME contribution.

Theoretical description of CME is also difficult: both $\m_A$ and $B$ contain large uncertainties. While their peak values are known to be set by respectively axial charge production in glasma phase \cite{Fukushima:2010vw,Mace:2016svc} and moving charge of spectators \cite{Skokov:2009qp}, their further evolution is model dependent. Different theoretical frameworks such as AVFD (anomalous viscous fluid dynamics) \cite{Jiang:2016wve,Shi:2017cpu,Hirono:2014oda,Shi:2019wzi}, chiral kinetic theory \cite{Huang:2017tsq,Sun:2016nig,Sun:2016mvh} and multiphase transport model \cite{Deng:2016knn,Ma:2011uma} have been employed to study the time evolution of axial/vector charges. All of these frameworks treat axial charge as an approximately conserved quantity in the absence of parallel electric and magnetic fields. However, it is also known that axial charge is not conserved due to gluon dynamics. In fact, it is the same origin for initial axial charge. In \cite{Lin:2018nxj}, three of us incorporated both fluctuation and dissipation of axial charge in the framework of stochastic hydrodynamics. It has been found that independent of the initial condition, the variance of axial charge always approaches thermodynamic limit in sufficient long time due to interplay of fluctuation and dissipation. In \cite{Lin:2018nxj}, we use the thermodynamic limit for the axial charge to model CME. While being model independent, the study misses an important fact: most charge separation occurs at very early stage of quark-gluon plasma, when both $\m_A$ and $B$ have not decayed appreciably. The purpose of this study is to incorporate the initial axial charge and investigate the coupled dynamics of axial and vector charge. In particular, we will give prediction for CME contribution for isobar collisions.

This paper is organized as follows: in Section 2, we generalize the stochastic hydrodynamics framework to include both axial and vector charge, which are coupled through anomalous effect in the presence of magnetic field. We will justify for phenomenological relevant magnetic field the back-reaction of vector charge to axial charge is negligible. In Section 3, we derive axial charge evolution with a non-vanishing initial value. The obtained axial charge is used for calculating charge separation. We will make prediction for CME in isobar collisions using AuAu collisions as a reference. We conclude and discuss future directions in Section 4.

\section{Stochastic Hydrodynamics for Axial and Vector Charges}

The stochastic hydrodynamic equations for axial charge in the absence of magnetic field have been written down in \cite{Lin:2018nxj}. In the presence of magnetic field, axial charge are coupled to vector charge through chiral magnetic effect and chiral separation effect. The full stochastic hydrodynamic equations for axial and vector charges are given by
\be\label{na_hydro}
\1\{\begin{split}
\nabla_\m J_A^\m&=-\frac{n_A}{\tau_{\cs}}-2\x_q \\
J_A^\m&=n_A u^\m+\l n_V eB^\m-\s TP^{\m\n}\nabla_\n\(\frac{n_A}{\chi_A T}\)+\x_A^\m,
\end{split}\2.
\ee
and
\be\label{nv_hydro}
\1\{\begin{split}
\nabla_\m J_V^\m&=0 \\
J_V^\m&=n_V u^\m+\l n_A eB^\m-\s TP^{\m\n}\nabla_\n\(\frac{n_V}{\chi_V T}\)+\x_V^\m,
\end{split}\2.
\ee
Here $n_A$ and $n_V$ are axial and vector charge density respectively. The axial current is not conserved due to topological configuration of gluons, which gives rise to the dissipative term $\sim\frac{n_A}{\t_\cs}$ and fluctuating noise term $\sim\x_q$. The constitutive equations for axial and vector current consist of co-moving term, anomalous mixing term, diffusive term and thermal noise term. $u^\m$ is the fluid velocity, which defines the projection operator $P^{\m\n}=g^{\m\n}+u^\m u^\n$ and the magnetic field in the fluid cell $B^\m=-\frac12\frac{\e^{\m\n\a\b}}{\sqrt{-g}}F_{\a\b}u_\n$.
The $\x_A$, $\x_V$ and $\x_q$ are taken to be Gaussian white noises:
\begin{align}
&\<\x_A^\m(x)\x_A^\n(x')\>=P^{\m\n}2\s_AT\frac{\dif^4(x-x')}{\sqrt{-g}}, \no
&\<\x_V^\m(x)\x_V^\n(x')\>=P^{\m\n}2\s_VT\frac{\dif^4(x-x')}{\sqrt{-g}}, \no
&\<\x_q(x)\x_q^\n(x')\>=\G_\cs\frac{\dif^4(x-x')}{\sqrt{-g}}, \no
&\<\x_A^\m(x)\x_q(x')\>=\<\x_V^\n(x)\x_q(x')\>=0,
\end{align}
with $\G_\cs$ being the Chern-Simon diffusion constant characterizing the magnitude of topological fluctuation.

For application to CME in heavy ion collisions, we fix the parameters as follows: we use free theory limit for axial and vector charge susceptibilities $\chi_A=\chi_V=\c=N_fN_cT^2/3$. The coefficient of the mixing term $\l$ are determined by chiral magnetic/separation effect as
$\l=\frac{1}{\chi}\frac{N_c}{2\pi^2}$. The quark mass effect on CSE can be neglected \cite{Lin:2018aon}. For three flavours, we have $\c=3T^2$ and $\l=\frac {1}{2\pi^2T^2}$. $\G_\cs$ is the Chern-Simon diffusion constant, for which we take from the extrapolated weak coupling results $\G_\cs=30\a_s^4T^4$ \cite{Moore:2010jd} with $\a_s=0.3$. The relaxation time of axial charge $\t_\cs$ is fixed by the Einstein relation as $\t_\cs=\frac{\c T}{2\G_\cs}$. $\s_A$ and $\s_V$ are conductivities for axial and vector current. We will not need their values in our analysis below.

The axial/vector charge is considered as perturbation in the background hydrodynamic flow. We will consider heavy ion collisions at top RHIC collision energy $\sqrt{s_{\text{NN}}}=200\gev$ and use Bjorken flow as the background.
In Milne coordinates $(\tau, \eta, x, y)$, the fluid velocity reads $u^\m=(1,0,0,0)$. We can show the total axial charge is conserved up-to mixing term and the topological fluctuation induced terms. To see that, we substitute the constitutive equation into the conservation equation in \eqref{na_hydro} and integrate over the volume $\int\t d\h d^2x_\perp=\int\sqrt{-g}d\h d^2x_\perp$. Using the identity $\nabla_\m V^\m=\frac{1}{\sqrt{-g}}\pd_\m\(\sqrt{-g}V^\m\)$ and dropping the boundary terms, we obtain
\begin{align}
\int d\h d^2x_\perp\(\pd_\t\(\t n_A\)-\pd_\t\(\s_ATP^{\t\n}\nabla_\n\(\frac{n_A}{\c T}\)\)+\pd_\t\(\t\x_A^\t\)\)=\int d\h d^2x_\perp\(-\frac{\t n_A}{\t_\cs}-2\t\x_q\).
\end{align}
Note that $P^{\m\n}u_\n=0$ and $\x_A^\m\x_A^\n\sim P^{\m\n}$, thus $P^{\t\n}=0$ and $\x_A^\t=0$. We then arrive at
\begin{align}\label{total}
\pd_\t N_A=-\frac{N_A}{\t_\cs}-\int d\h d^2x_\perp2\t\x_q,
\end{align}
with $N_A=\int\t d\h d^2x_\perp$. The absence of diffusive term, thermal noise term and mixing term is consistent with the picture that these terms only lead to redistribution of axial charge. The counterpart for vector charge is simpler: $\pd_\t N_V=0$ because vector charge is strictly conserved.


\subsection{The Back-Reaction from the Vector Current}

We will assume the following distribution of axial charge: the initial axial charge created by chromo flux tube is homogeneous in transverse plane. The boost invariant Bjorken expansion maintains a homogeneous distribution in longitudinal direction. The homogeneous axial charge gives rise to charge separation via CME. 
This simplified picture is modified by three effects: diffusion, thermal noise and CSE. The thermal noise and diffusion correspond to fluctuation and dissipation of charge, which bring the charge to equilibrium. The CSE is not balanced by other effect. We show now its effect is sub-leading.

Let us compare the axial charge $n_A$ and the CSE modification $\sim\l B n_V$. Since $\chi_A=\chi_V$, it is equivalent to compare $\m_A$ and $\l B\m_V$. Since $B$ drops quickly with time, the CSE effect is maximized at initial time. We estimate the initial $n_A$ following \cite{Jiang:2016wve} as
\begin{align}\label{na0}
\sqrt{\lag n_A(\t_0)^2\rag}\simeq \frac{Q_s^4(\pi \r_{\text{tube}}^2\t_0)\sqrt{N_{\text{coll}}}}{16\pi^2S_\perp},
\end{align}
where $Q_s$ is the saturation scale and $\r_{\text{tube}}\simeq 1\text{fm}$ is the width of the flux tube. $\t_0$ is the initial proper time. For AuAu collisions, we take $Q_s\simeq 1\gev$ and $\t_0=0.6\fm$. The number of binary collisions $N_{\text{coll}}$ and the transeverse overlap area $S_\perp$ and calculated using Monte Carlo Glauber model \cite{Miller:2007ri,Alver:2008aq,Loizides:2014vua,Loizides:2017ack} with the centrality dependence listed in Table \ref{tab:MCG}. 

\begin{table}
\caption{\label{tab:MCG} Geometrical quantities from MC Glauber model for Au, Ru and Zr. $N_{\text{coll}}$, $S_\perp$ and $L_\perp$ are number of binary collisions, transverse overlap area, and the width of the participants' region along the cross-line between the transverse plane and the reaction plane. $S_\perp$ is taken to be the projection of the nucleon-nucleon cross-section $\s_\text{NN}$ onto the transverse plane \cite{Abelev:2008ab}, and $L_\perp$ is calculated through the same algorithm as $S_\perp$. $10k$ events are run to generate the datas. Averages are done using the impact parameter $b$ as the weight factor.}
\begin{tabular}{ccccccccccc}
\hline
\text{Centrality}& 0-5\%& 5-10\%& 10-20\% & 20-30\% &30-40\%& 40-50\%& 50-60\%& 60-70\% & 70-80\% \\
\hline
&&&&\textbf{Au}\\
\hline
$N_\text{coll}$& 1049.8 &843.9 &594.8 &369.1 &217.4 &121.6 &62.2 &29.2 &12.7\\
\hline
$S_\perp(\fm^2)$& 147.9 &128.9& 106.1 & 83.0 &64.8 &49.7 &36.6 &25.5 & 16.2\\
\hline
$L_\perp(\fm)$& 13.2 & 11.9& 10.3 & 8.6& 7.3 &6.1 & 5.0 & 4.1 & 3.1\\
\hline
&&&&\textbf{Ru}\\
\hline
$N_\text{coll}$& 387.5&316.3&228.9&146.6&90.9&53.6&30.0&15.8&8.1\\
\hline
$S_\perp(\fm^2)$& 92.5&81.6&67.8&53.8&42.3&32.8&24.7&17.6&12.1\\
\hline
$L_\perp(\fm)$&10.5&9.5&8.3&7.0&6.0&5.1&4.3&3.5&2.9\\
\hline
&&&&\textbf{Zr} \\
\hline
$N_\text{coll}$& 395.6&322.5&232.1&149.0&91.8&54.0&30.1&15.7&8.0\\
\hline
$S_\perp(\fm^2)$& 91.3&80.5&67.0&53.1&41.8&32.5&24.4&17.4&11.9\\
\hline
$L_\perp(\fm)$& 10.4&9.4&8.2&7.0&6.0&5.0&4.2&3.5&2.9\\
\hline
\end{tabular}
\end{table}

The initial temperature is taken as $T_0=350\mev$. These combined give $\m_A\simeq 36\mev$ with very weak centrality dependence. On the other hand, $\m_V$ is estimated from \cite{BraunMunzinger:2001as}
\begin{align}
\m_B(s)\simeq\frac{a}{1+\sqrt{s}/b},
\end{align}
with $a\simeq1.27 \gev$ and $b\simeq4.3\gev$. At $\sqrt{s_\text{NN}}=200\gev$, $\m_B\simeq 27\mev$, corresponding to $\m_V\simeq 9\mev$. Taking peak value of $B\simeq 10m_\pi^2$, we find $\l eB\m_V/\m_A\simeq 3\%$. Since the magnetic field decays rapidly with time, a more realistic estimation for the back-reaction is to use time-averaged magentic field. Assuming the following functional form of magnetic field \cite{Yin:2015fca, Yee:2013cya},
\begin{align}\label{bdecay}
eB(\t)=\frac{eB_0}{1+(\t/\t_B)^2},
\end{align}
and averaging between initial time $\t_0=0.6\text{fm}$ and freeze-out time $\t=7\text{fm}$, we obtain $\l {eB}_{avg}\m_V/\m_A\simeq 1\%$ for $\t_B=2\text{fm}$ and $\l {eB}_{avg}\m_V/\m_A\simeq 0.4\%$ for $\t_B=1\text{fm}$. Therefore we can safely neglect the CSE effect on axial charge redistribution. Similar analysis shows the same is true for isobar collisions.

\subsection{The Evolution of the Axial Chemical Potential}

Since the back-reaction from vector charge is negligible, we can simply trace the evolution of total axial charge and use it to determine the average $\m_A$ for CME phenomenology.
In \cite{Lin:2018nxj}, we have derived the hydrodynamic evolution of the total axial charge with an initial value. It is given by
\be
\lag N_A(\t)^2\rag=\lag N_A(\t_0)^2\rag e^{3\(1-\(\frac{\t}{\t_0}\)^{2/3}\)\(\frac{\t_0}{\t_{\cs0}}\)}+\int \dif\et \dif^2x_\perp 2\G_0\t_0\t_{\cs0}\(1-e^{3\(1-\(\frac{\t}{\t_0}\)^{2/3}\)\(\frac{\t_0}{\t_{\cs0}}\)}\).
\ee
The initial condition for Au+Au collisions at $\sqrt{s_{\text{NN}}}=200\gev$ has been discussed in the previously subsection. The counterpart for isobars scales accordingly. We adopt the scaling of $Q_s$ with system size in \cite{Kharzeev:2000ph} and the initial time for Bjorken hydrodynamics in \cite{Basar:2013hea}. The freeze-out time is determined by the same freeze-out temperature $T_f=154\mev$. We list the scalings as follows,
\be\label{eq:para}\begin{split}
Q_s\sim A^{\frac16}, &\qquad T_0\sim Q_s\sim A^{\frac16}, \\
\t_0\sim 1/{Q_s} \sim A^{-\frac16}, &\qquad \t_f\sim A^{\frac13}.
\end{split}\ee
The axial chemical potential is calculated using the average axial charge
\be\label{ma_full}
\m_A(\t)=\frac{\sqrt{\lag n_A(\t)^2\rag}}{\c(\t)}=\frac{\sqrt{\lag N_A(\t)^2\rag}}{V(\t)\ \c(\t)},
\ee
with $V(\t)=S_\perp\t\D\et$ being the total volume. The rapidity span is taken to be $|\h|<2$ with $\D\h=4$. It determines the axial chemical potential as
\be\label{muatau}
\m_A(\t)=\m_{A0}\1(\frac{\t}{\t_0}\2)^{-\frac13} \sqrt{e^{3\(1-\(\frac{\t}{\t_0}\)^{2/3}\)\(\frac{\t_0}{\t_{\cs0}}\)}+\frac{3T_0^3}{\t_0\ \D\et\ S_\perp\ n_{A0}^2}\1[1-e^{3\(1-\(\frac{\t}{\t_0}\)^{2/3}\)\(\frac{\t_0}{\t_{\cs0}}\)}\2]},
\ee
where the square root factor is a modification to the simple $\t^{-1/3}$ dependence when relaxation of axial charge is ignored. The initial axial chemical potential is determined by the initial axial charge density $n_{A0}$ given in equation~\eqref{na0} via $\mu_{A0}=\frac{n_{A0}}{\chi_0}=\frac{n_{A0}}{3T_0^2}$.

Then we determine the scalings of the initial axial charge density and chemical potential. From the empirical scaling for AuAu collisions \cite{Miller:2007ri,Abelev:2008ab} in Glauber model,
\be\label{SN}
S_\perp \sim N_\text{part}^{\frac23}, \qquad N_\text{coll} \sim N_\text{part}^{\frac43},
\ee
where $N_\text{part}$ is the number of participant nucleons, we have
\be
S_\perp \sim \sqrt{N_\text{coll}}.
\ee
Thus from \eqref{na0} $n_{A0}$ has only weak centrality dependence. The system size dependence of $n_{A0}$ and $\m_{A0}$ can be easily obtained using \eqref{eq:para}
\be\label{nmuA}
n_{A0} \sim A^{\frac12}, \qquad \m_{A0} \sim A^{\frac16}.
\ee
The centrality dependence of initial chemical potential $\m_{A0}$ for Au and isobars are listed in Table \ref{tab:cen_m50}. Indeed we see weak centrality dependence for AuAu and slightly enhanced dependence for Ru and Zu due to deviation from the empirical scaling \eqref{SN}. The system size dependence \eqref{nmuA} is approximately consistent with Table \ref{tab:cen_m50}.

\begin{table}[h]
\caption{\label{tab:cen_m50}The centrality dependence of $\m_{A0}(\mev)$.}
\begin{tabular}{cccccccccc}
\hline
\text{Centrality}& 0-5\%& 5-10\%& 10-20\% & 20-30\% &30-40\%& 40-50\%& 50-60\%& 60-70\% & 70-80\% \\
\hline
$\text{Au}$& 36.11& 37.15& 37.90& 38.14& 37.53& 36.56& 35.49& \
34.97& 36.19\\
\hline
$\text{Ru}$& 31.13& 31.89& 32.63& 32.93& 32.99& 32.63& 32.45& \
33.06& 34.55\\
\hline
$\text{Zr}$& 31.85& 32.62& 33.29& 33.62& 33.51& 33.08& 32.89& \
33.35& 34.84\\
\hline
\end{tabular}
\end{table}

\section{Chiral Magnetic Effect in Isobar Collisions}

\subsection{The Effective Electrical Chemical Potential for Isobars}

Now we can calculate the chiral magnetic current using \eqref{cme}, whose time integral gives the total charge separation
\be\label{eq:qv}
Q_e=\int_{\t_0}^{\t_f} \dif\t\ \t \dif\et L_\perp\ C_e\m_A\ eB = C_e\D\et\ L_\perp\int_{\t_0}^{\t_f} \dif\t \ \t \m_A(\t)\ eB(\t),
\ee
where $C_e=\sum_f\frac{q_f^2}{e}\frac{N_c}{2\pi^2}$ and $L_\perp$ is the width of the participants' region along the cross-line between the transverse plane and the reaction plane, sampled from the MC Glauber Model, see Table \ref{tab:MCG}.  Hence $\int \t d\et L_\perp$ represents the area that the CME current penetrates in the reaction plane. We integrate it from initial thermalization time $\t_0$ to freeze-out time $\t_f$, with their values determined in \eqref{eq:para}. The effective electric chemical potential is then induced by the total electric charge asymmetry as,
\be\label{eq:mu_v}
\m_e(\t_f)=\frac{Q_e}{V_f\ \chi_e(\t_f)}=\frac{3 L_\perp}{\p^2\  eS_\perp  \t_f\ T_f^2} \int_{\t_0}^{\t_f} \dif\t \t \m_A(\t)\ eB(\t),
\ee
where $V_f=S_\perp \t_f \D\et/2$ and $\chi_e(\t_f)=\frac{1}{3}\sum_fq_f^2N_cTf^2$ denoting the volume of QGP above the reaction plane and the electric charge susceptibility at freeze-out time.

\begin{figure}[h]
\centering
\includegraphics[scale=0.8,angle=0]{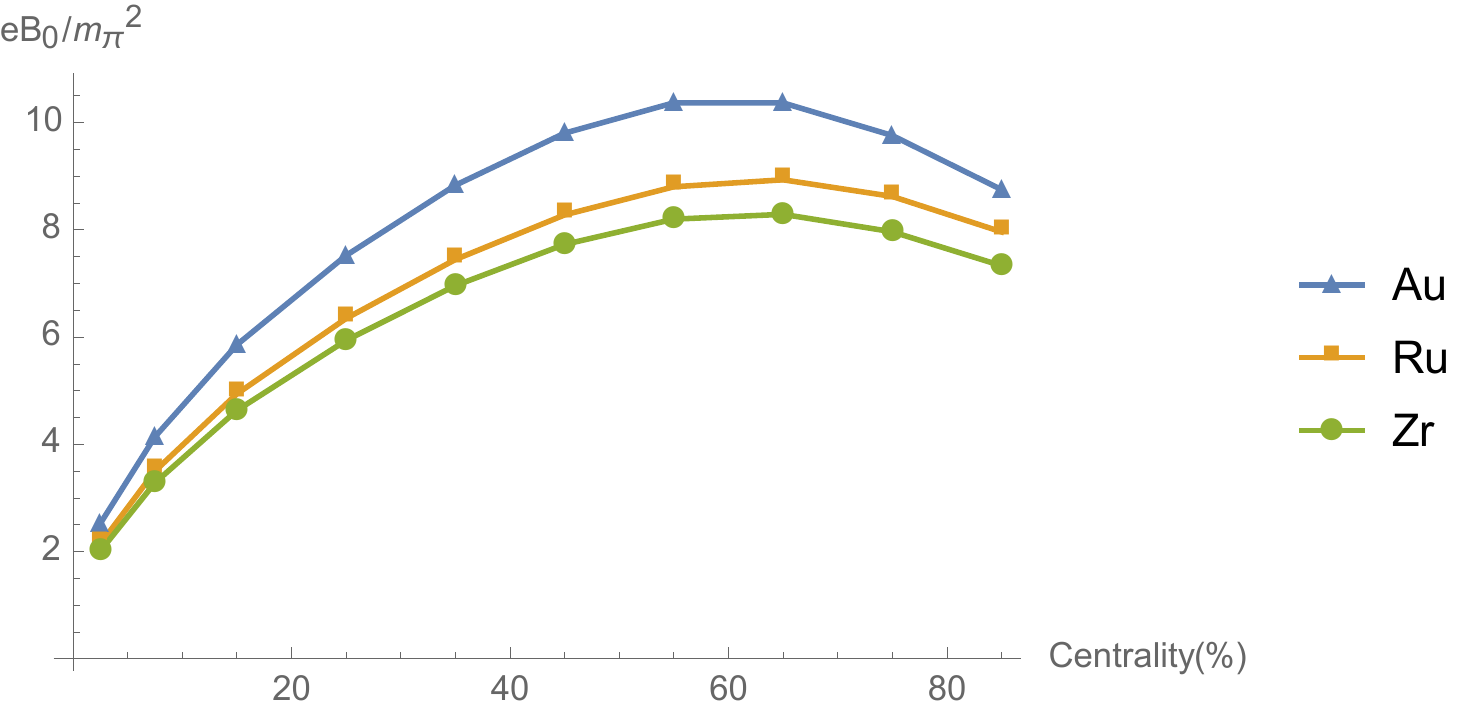}
\caption{\label{Byb} Centrality dependence of the event-averaged magnetic field oriented out of the reaction plane, with triangles for Au, squares for Ru and circles for Zr.}
\end{figure}

The magnetic field in the lab frame is calculated from the Li\'enard-Wiechert potentials as
\be\label{eq:lw}
e{\bf B}(t, {\bf r})=\frac{e^2}{4\pi}\int \dif {\bf r'}^3  \rho_Z(r')\ \frac{1-{\bf v}^2}{\1[{\bf R}^2-({\bf R}\times {\bf v})^2\2]^{3/2}}{\bf v}\times {\bf R}\
\ee
where ${\bf R}={\bf r}-{\bf r'}(t)$ is the vector pointing from the proton position ${\bf r}(t)$ at time $t$ to the position ${\bf r}$ of the field point. ${\bf v}$ is the velocity of the protons, chosen to be $v^2=1-(2m_N/\sqrt{s})^2$, where $\sqrt{s}/2$ is the energy for each nucleon in the center-of-mass frame, and $m_N$ is the mass of  the nucleon. The impact parameter vector is set to be along the $x$-axis so that the $x-z$ plane would serve as the reaction plane and $x-y$ as the transverse plane. We sample the positions of protons in a nucleus in the rest frame by the Woods-Saxon distribution, 
\be
\rho_Z(r')\propto \frac{1}{1+\exp{\1(\frac{r'-R_0}a\2)}},
\ee
where $R_0=6.38\text{fm}$ and $a=0.535\text{fm}$ for Au, and $R_0=5.085 \text{fm}$ and $5.020 \text{fm}$ for Ru and Zr respectively, and $a = 0.46 \text{fm}$ for both isobars. 
The homogeneous and boost invariant power-decaying form of the magnetic field is assumed by equantion~\eqref{bdecay}
with the peak value $eB_0$ set by equation~\eqref{eq:lw} at $t={\bf r}=0$ along the $y$-axis.
%
Dependence on nucleus shape discussed in \cite{Xu:2017zcn} is not included in our analysis. As a result, the centrality dependence of $eB_0$ for Au, Ru and Zr are shown in Figure~\ref{Byb}. We see that the magnitude of the magnetic field is suggested by the proton numbers of the corresponding nucleaus, and the difference between isobars is indicated as $\sim 10\%$.

The characteristic decay time of the magnetic field $\t_B$ has a large uncertainty in different models \cite{McLerran:2013hla,Gursoy:2014aka,Tuchin:2015oka}, we treat it as a fitting parameter and fix it by matching the CME signal for AuAu collisions calculated in our model to the flow-excluded charge separation measurement by the STAR collaboration at $\sqrt{s_\text{NN}}=200 \gev$ \cite{Adamczyk:2014mzf}, see Section~\ref{expdata}. This gives $\t_B= 1.65\text{fm}$. We will assume the same $\t_B$ for isobars at the same collision energy, and use our model to make predictions for CME signals for Ru and Zr.

\begin{figure}[h]
\centering
\includegraphics[scale=0.8,angle=0]{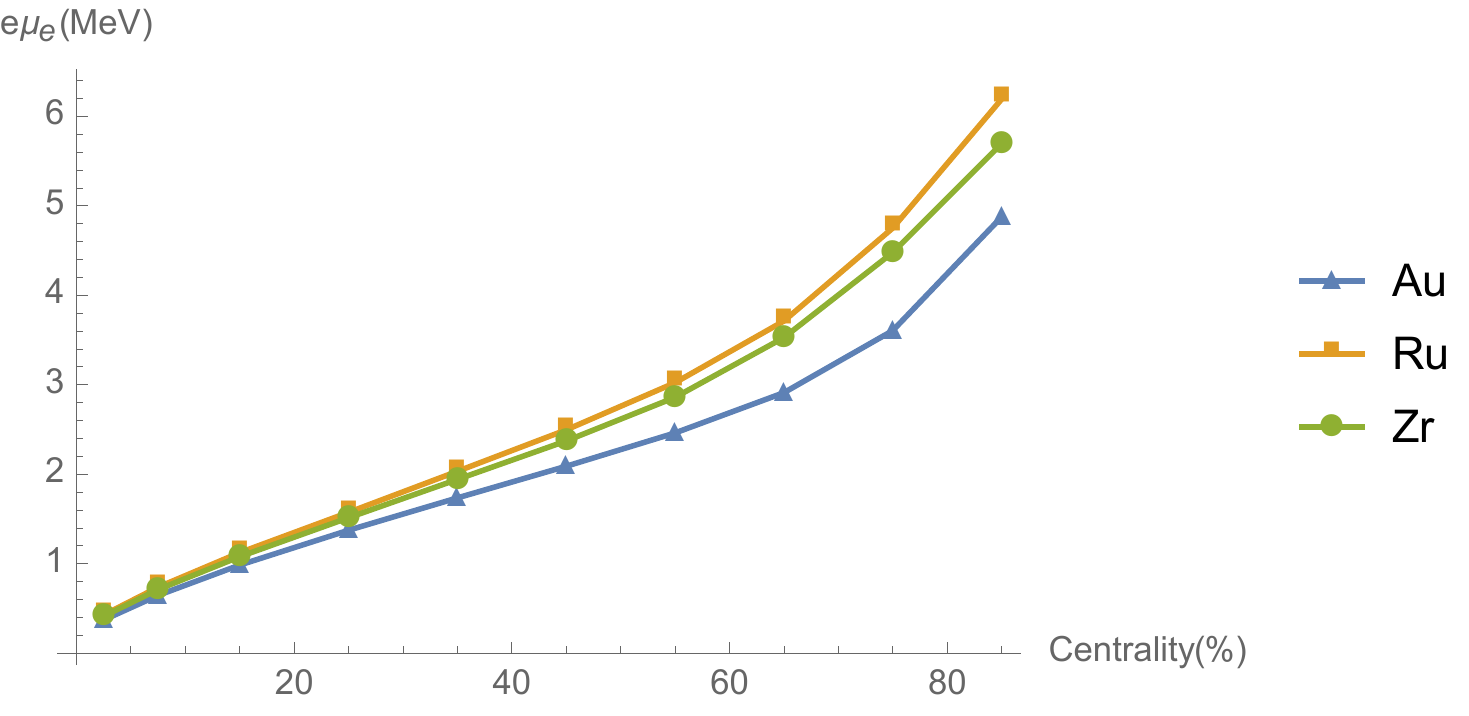}
\caption{\label{emuv} Centrality dependence of the event-averaged electric chemical potentials induced by the chiral magnetic effect, with triangles for Au, squares for Ru and circles for Zr.}
\end{figure}

Finally, we obtain a $e\m_e$ for different centralities in Figure~\ref{emuv}. Despite the system of AuAu having larger $\m_{A0}$ and $eB$, it gives smaller $e\m_e$ than the systems of Ru and Zr. This is due to the larger volume factor in \eqref{eq:mu_v}.
We will obtain the scaling in the following subsection. 

\subsection{The scaling relationship of the electrical chemical potential for different heavy-ions}\label{scaling}
To determine the scalings of the magnitude of the electric chemical potential for different heavy-ions, we plug equation~\eqref{bdecay} and \eqref{muatau} into equation~\eqref{eq:mu_v} and sort it into several blocks as
\begin{align}
\m_e(\t_f)&=\frac{3}{\p^2\ T_f^2}\ \frac{L_\perp B_0}{ S_\perp}\ \frac{1}{\t_f}\ \int_{\t_0}^{\t_f}  \dif\t\ \frac{\t }{1+(\t/\t_B)^2}\ \m_{A0}\1(\frac{\t}{\t_0}\2)^{-\frac13} \times\nonumber\\
&\sqrt{e^{3\(1-\(\frac{\t}{\t_0}\)^{2/3}\)\(\frac{\t_0}{\t_{\cs0}}\)}+\frac{3T_0^3}{\t_0\ \D\et\ S_\perp\ n_{A0}^2}\1[1-e^{3\(1-\(\frac{\t}{\t_0}\)^{2/3}\)\(\frac{\t_0}{\t_{\cs0}}\)}\2]}.
\end{align}
%
The first block $\frac{3}{\p^2\ T_f^2}$ holds the same for three types of nucleus. The second block $ \frac{L_\perp B_0}{ S_\perp}$ is determined entirely from the geometry of the nuclei, i.e, the distribution of nucleons. The third block $\frac{1}{\t_f}\ \int_{\t_0}^{\t_f}$ accounting for the integral average scales as $\frac{\t_f-\t_0}{\t_f}\sim 1$. The fourth block $\m_{A0}\1(\frac{\t}{\t_0}\2)^{-\frac13}$ is determined by the initial condition from the glasma, for which we already discussed in Section 2. The square root factor accounts for the damping and fluctuation in our stochastic model.

We first determine the scaling of the geometrical term $\frac{L_\perp B_0}{ S_\perp}$. Throughout the following analysis, the empirical proportionality relationship $R_0\sim A^{1/3}$ is implied. For $L_\perp$ and $S_\perp$, the geometrical property from the Glauber model is straightforward,
\be
S_\perp\sim R_0^2 \sim A^{2/3}, \qquad L_\perp\sim R_0\sim A^{1/3},
\ee
which is also in agreement with equation~\eqref{SN}, if we assume the number of participants scales with the volume $N_\text{part}\sim R^3 \sim A$.

To analyze the magnetic field, we have to know its dependence on the centrality. Note that equation~\eqref{eq:lw} is the dependence on the impact parameter, but at a given centrality, the averaged impact parameter is different for three types of nuclei. Since we are comparing the signal in each fixed centralities, we have to know how the averaged impact parameter scales for different nuclei in each centrality.

Following from \cite{Miller:2007ri}, the distribution of the total cross section $\s_\text{tot}$ holds well for $b<2R_0$ ,
\be
\frac{\dif \s_\text{tot}}{\dif b}\simeq 2\pi b,
\ee
thus the total cross section scales as, 
\be
\s_\text{tot}\sim \int^{R_0} b\dif b\sim R_0^2\sim A^{2/3},
\ee
which is a reasonable scaling in term of dimensions.
Then quoting from \cite{Broniowski:2001ei}, the following geometric relation between centrality $c$ and the impact parameter $b$ also holds to a very high precision for $b<2R_0$,
\be
b(c)\simeq\sqrt{\frac{c\cdot \s_\text{tot}}{\pi}},
\ee
thus for a given centrality $c$, the average impact parameter for different nucleus scales with 
\be
b(c)\sim \s_\text{tot}^{1/2} \sim A^{1/3}.
\ee
To proceed to determine the scaling of the magnetic field, we take the multiple-pole expansion of equation~\eqref{eq:lw} and 
treat the monopole as our scaling of the magnetic field for different nucleus at a given centrality $c$, thus it is given by
\be
B_0(c) \sim Z/b(c)^2 \sim ZA^{-2/3}.
\ee
Therefore the geometrical combination block scales as
\be
\frac{L_\perp B_0}{S_\perp}\sim \frac Z A.
\ee

Next, we look at the chemical potential block, without damping and fluctuation effect. The scaling of the initial chemical potential is already discussed in Section 2, it's $\m_{A0}\sim A^{1/6}$, but considering the volume expansion which contains $\t_0$, it scales as
\be
\m_{A0}\1(\frac{\t}{\t_0}\2)^{-\frac13}\sim A^{1/9}.
\ee

Lastly, the most ambiguous block is the square root factor accounting for the damping and fluctuation effect. From the above analysis, the scaling of the fluctuation is set by
\be
\frac{3T_0^3}{\t_0\ \D\et\ S_\perp\ n_{A0}^2} \sim A^{-1}.
\ee
But fluctuation is generally small compared to initial contribution from the glasma, so if we neglect it, the square root factor just scales with $1$. Incorporating contribution from both of them, we may write the scaling of the square root factor as $A^{-\zeta}$, with $0<\zeta<\frac12$.\\

Putting all the above together, we have the scaling of the electric chemical potential as
\be
\m_e(\t_f) \sim \(\frac Z A\) A^{\frac19} A^{-\zeta} = Z A^{-(\zeta+\frac89)},
\ee
with $0<\zeta<\frac12$. When we consider only the CME from the initial condition, $\zeta=0$; when we consider only the fluctuation effect $\zeta=\frac12$; otherwise, $\zeta$ lies between them. Our numerical datas for Au and isobars suggest a rough value of  $\zeta\simeq\frac14$; but note that there're deviations in each centrality mainly due to our simplified scaling of the magnetic field using the monopole.

\subsection{The CME signal to be compared in experiments}\label{expdata}
To proceed, we would firstly need Cooper-Frye freeze-out procedure~\cite{Cooper:1974mv} to give the spectrum of the single particle distribution as,
\be
E\frac{\dif N}{\dif^3 p}=\frac{g}{(2\pi)^3}\int p^\mu \dif^3\sigma_\mu f(x,p),
\ee
where $g$ is the degeneracy factor, taken to be $1$ for each species of mesons ($K^\pm$, $\pi^\pm$) produced in QGP respectively. The 4-momentum of the particle and the Bjorken spacetime 4-velocity are given by
\be
p^\m=(m_\perp \cosh y,\ p_\perp, m_\perp \sinh y), \qquad  u^\m=(\cosh \et, 0,0,\sinh\et),
\ee
with $m_\perp=\sqrt{p_\perp^2+m^2}$. Note that $y$ is the particle rapidity and $\et$ is the spacetime rapidity.
Thus we could expand the Cooper-Frye formula as
\be\label{cf}
\frac{\dif N}{\dif \phi \dif y p_\perp \dif p_\perp}=\frac{g}{(2\pi)^3}\int \tau_f\dif\et\dif^2x\ m_\perp \cosh(\eta-y)f(x,p).
\ee
The phase-space distribution of the $i$-th particle species at freeze-out time is given in Boltzmann approximation as, 
\be
f_i(x,p)=e^{(p_\m u^\m \pm e\m_e(\t_f)+\m_i)/T_f},
\ee 
where $\pm\m_e(\t_f)$ is the positive or negative electric chemical potential at freeze-out time caused by CME, see Figure~\ref{emuv}, which is much smaller than the freeze-out temperature $T_f\simeq 154\mev$~\cite{Teaney:2002aj}, and $\mu_i$ is the chemical potential for $i$-th species, here we consider only pions and kaons in our
calculations with respect to heavy-ion collisions, with $\m_\p\simeq 80\mev$ for pions and $\m_K\simeq 180\mev$ for kaons. Thus we can approximate the distribution to the lowest order in $\m_e$ as
\be
\d f_i(x,p)=f_i(\mu_e=0)\ \frac{\pm e\mu_e(\t_f)}{T_f},
\ee
this leads to the azimuthal distribution of the $i$th positive or negative charged particle $N_\pm^i$ created from CME as
\begin{align}
\label{eq:dndp0}
\d \frac{\dif N_\pm^i}{\dif \phi}=\frac{g_i\ S_\perp}{(2\pi)^3}\int\!  \dif m_\perp m_\perp^2 
\int\! \t_f \dif y \dif \et\  \cosh(\et-y) f_i(\mu_e=0)\ \frac{\pm e\mu_e(\t_f)}{T_f},
\end{align}
where we used the fact that $p_\perp\dif p_\perp =m_\perp\dif m_\perp$. The lower bound of $m_\perp$ integration being the rest mass of corresponding meson. The integration domain for particle-rapidity should be taken according to experiments as $|y|<1$, and the space-time rapidity as $|\et|<2$. Note that the sign difference on the RHS of the above equation, the charge asymmetry of the particle distribution is due to CME. Since the magnetic field points to the upper half of the QGP region from the lower half across the reaction plane, positive charge accumulates in the above and negative in the below, thus $\m_e$ changes sign cross the reaction plane.
Similarly, the multiplicity of charged particles from the background is obtained consistently from equation~\eqref{cf} as
\begin{align}\label{bgi}
\frac{\dif N_\pm^i}{\dif \phi}=\frac{g_i\ S_\perp}{(2\pi)^3}\int\!  \dif m_\perp m_\perp^2 
\int\! \t_f \dif y\dif\et  \cosh(\et-y) f_i(\m_e=0),
\end{align}
where there shows no sign difference between positive and negative charges, indicating that the background is electric-neutral.

To get the total charged particle multiplicity from CME $\D_\pm$ and from the neutral background $N_\pm^{bg}$, the index $i$ should be summed over different species, thus we define
\be\label{Del}
\D_\pm\equiv \sum_i \d N_\pm^i, \qquad  N^{bg}_\pm\equiv \sum_i  N^i_\pm,
\ee
where again $\pm$ denotes positive or negative charge. Note that since we assume the whole QGP is electric-neutral, the fluctuation of the electric chemical potential is averaged to be zero, $\lag \m_e(\t_f)\rag =0$, but the two-point correlation is taken to be the square of the electric chemical potential itself, $\lag \m_e(\t_f)^2\rag\simeq \m_e(\t_f)^2$. Also note that our electric chemical potential $\m_e$ calculated in Section 2 is an effective quantity, it's not $\et$-dependent and decouples in the integrals. Then from equations~\eqref{eq:dndp0}, \eqref{bgi} and \eqref{Del}, denoting $\a,\b=\pm$ and $\s_\pm=\pm 1$, we have the following average and proportionality relations as
\be\label{ratio}
\lag \D_\a\rag=0, \qquad \frac{\lag \D_\a\D_\b\rag}{\< N^{bg}_\a\>\< N^{bg}_\b\>}\simeq\s_\a\s_\b\frac{(e\m_e(\t_f))^2}{T_f^2}.
\ee
The average relation on the left is interpreted straightforward as the conservation of electric charge.
The proportionality relation on the right is a measurement of the asymmetry. The CME induced term $\D_\pm$ is treated as a perturbation to the electric-neutral background as heat bath with temperature $T_f$. 

To move on, we analyze the background angular distribution $d\lag N_\pm\rag /d\phi$, which reflects the charge-independent evolution of the medium determined by the event-by-event fluctuating initial state. In this point, we take the Fourier expansion of the background angular distribution as
\be
\frac{\dif \lag N^{bg}_\pm\rag }{\dif\phi} = \frac{\lag N^{bg}_\pm\rag}{2\pi} 
\left[1+2\sum_{n=1} v_n \cos n(\phi-\Psi_{n})\right] ,
\ee
where $\Psi_n$ indicates the participant plane angle of order $n$. Note that we have dropped the sine term in the Fourier decomposition due to the fact that the distribution is symmetric about the participant plane. The coefficient $v_n$ is defined as the $n$th order harmonic flow. Typically, the directed flow $v_1$ is generally chosen to be $0$ if the distribution is measured in a symmetric rapidity region \cite{Bzdak:2012ia,Voloshin:2004vk}, thus in the following calculation we only kept the next leading term from the elliptic flow $v_2$.

To proceed, we assume the following ansatz \cite{Kharzeev:2007jp} for the total generated charged single-particle
spectrum originated from both the background and the CME,
\be\label{decomp}
\frac{\dif N_\pm}{\dif\phi}= \frac{\dif \lag N^{bg}_\pm\rag }{\dif\phi}+ \frac1 4 \D_\pm \sin(\phi-\Psi_{RP}),
\ee
where the form of the CME-induced term is proportional to $\sin(\phi-\Psi_{RP})$ owing to the symmetry of the distribution about the magnetic field, which is perpendicular to the reaction plane, and the factor $1/4$ is consistent with our definition \eqref{Del}.


Different from our previous work \cite{Lin:2018nxj}, we choose our correlated two-particle spectrum not just as a product of the single spectrum, but also including an underlying correlation term proposed in \cite{Bzdak:2012ia} as
\be
\rho(\phi_1,\phi_2)=\left\langle\frac{\dif N^\alpha}{\dif \phi_1^\alpha}\frac{ \dif N^\beta}{\dif \phi_2^\beta}\right\rangle\bigg[1+\sum_{n=0}^{\infty} a_n \cos n(\phi_1-\phi_2)\bigg],
\ee
with $\a,\,\b=\pm$. The cosine correlation term is reaction-plane-insensitive. Here we only take the leading term $a_1$ into consideration (with normalization leading to $a_0=0$).

\begin{figure}[h]
\centering
\includegraphics[scale=0.8,angle=0]{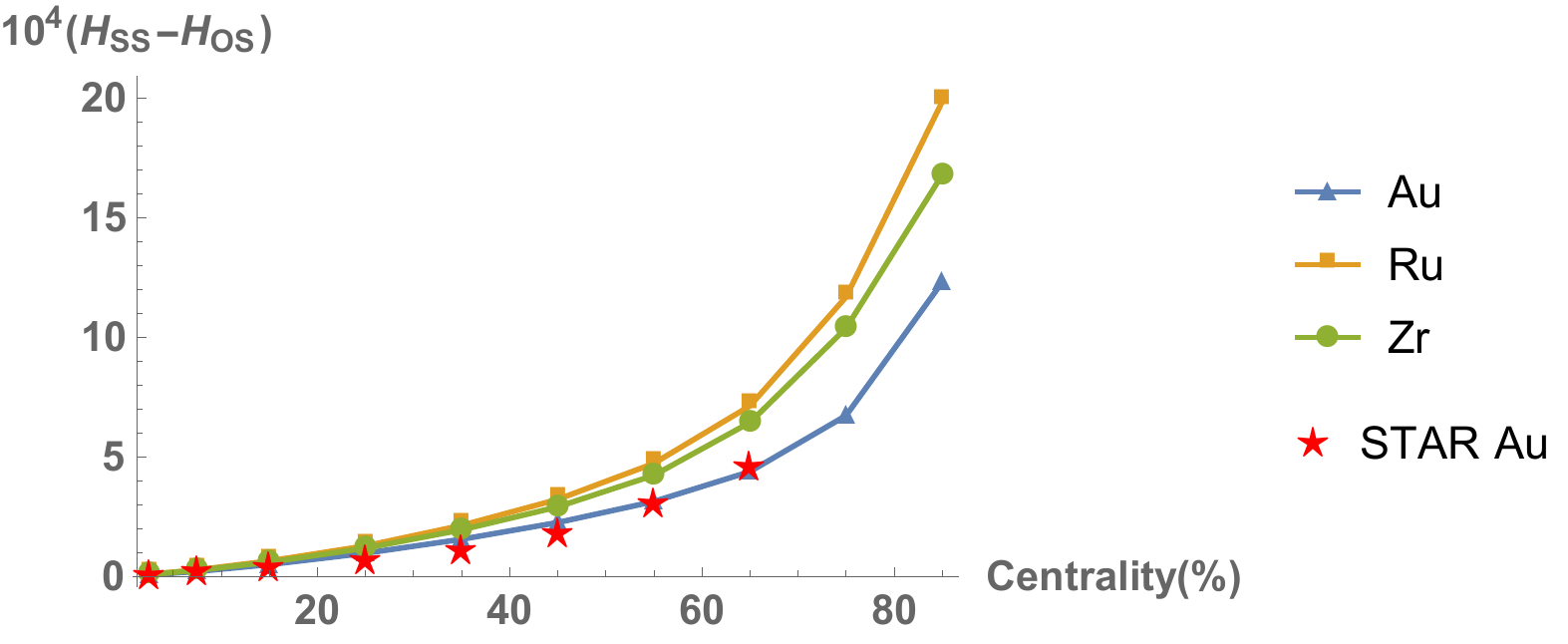}
\caption{\label{CME} Centrality dependence of the CME signal from our stochastic model for AuAu and isobaric collision at $\sqrt{s_\text{NN}}=200\gev$, with triangles for Au, squares for Ru and circles for Zr. We also list the datas for AuAu collisions at $\sqrt{s_\text{NN}}=200\gev$, extracted from STAR \cite{Abelev:2009ad, Agakishiev:2011eq}, with pentacles, for comparison.}
\end{figure}

With all of these, the two types of the two particle correlations $\gamma$ and $\delta$, measured in the heavy-ion collision experiments are given as
\be\1\{\begin{split}
\gamma_{\alpha\beta}&=\<\cos(\phi_1^\alpha+\phi_2^\beta-2\Psi_{\text{RP}})\>\\
\delta_{\alpha\beta}&=\<\cos(\phi_1^\alpha-\phi_2^\beta)\>,
\end{split}\2.\ee
where the average $\<\cos\varphi\>$ of the angle $\varphi=(\phi_1^\alpha+\phi_2^\beta-2\Psi_{\text{RP}})$ or $(\phi_1^\alpha-\phi_2^\beta)$ is taken over events, i.e, integrated over $\phi_1$ and $\phi_2$ as 
\be
\<\cos\varphi\>=\frac{\int\rho(\phi_1,\phi_2)\ \cos \varphi\ \dif\phi_1^\alpha \dif\phi_2^\beta}{\int\rho(\phi_1,\phi_2)\ \dif\phi_1^\alpha \dif\phi_2^\beta}.
\ee
This will result in 
\be\1\{\begin{split}
\gamma_{\alpha\beta}&
=\langle v_2 a_1 \cos2(\Psi_2-\Psi_{RP})\rangle-\frac{\pi^2}{16}\frac{\langle \Delta_\alpha \Delta_\beta\rangle }{\langle N^{bg}_\alpha\rangle \langle N^{bg}_\beta\rangle}\\
\delta_{\alpha\beta}&
=\langle \frac{a_1}{2}(1+v_2^2)\rangle+\frac{\p^2}{16}\frac{\langle \Delta_\alpha \Delta_\beta\rangle }{\langle N^{bg}_\alpha\rangle \langle N^{bg}_\beta\rangle}.
\end{split}\2.\ee
These forms of $\gamma$ and $\d$ correlators are consistent with the proposal in \cite{Adamczyk:2014mzf,Bzdak:2012ia}:
\be\label{gd}
\1\{\begin{split}
\gamma_{\a\b}&= \kappa v_2F_{\a\b}-H_{\a\b}\\
\delta_{\a\b}&= F_{\a\b}+H_{\a\b},
\end{split}\2.
\ee
with $F_{\a\b}$ denoting the background and $H_{\a\b}$ denoting the CME contribution, and $\kappa$ is an undetermined factor ranging from $1$ to $2$. Therefore, by matching the above sets of equations and using equation~\eqref{ratio}, we claim that the CME signal takes the following form
\be
H_{\a\b}=\frac{\p^2}{16}\frac{\langle \Delta_\alpha \Delta_\beta\rangle }{\langle N^{bg}_\alpha\rangle \langle N^{bg}_\beta\rangle}\simeq\s_\a\s_\b\ \frac{\p^2}{16}\frac{(e\m_e(\t_f))^2}{T_f^2}.
\ee
The difference between the same charge correlation $H_{SS}$ and the opposite charge correlation $H_{OS}$ is thus expressed as
\be
(H_{SS}-H_{OS})\simeq2\cdot\frac{\p^2}{16}\frac{(e\m_e(\t_f))^2}{T_f^2}.
\ee
The centrality dependence of $10^4\1(H_{SS}-H_{OS}\2)$ for Au and isobars are shown in Figure~\ref{CME}.
We also plot the signal for AuAu collision at $200\gev$ with datas extracted from STAR, by solving equation~\eqref{gd} as
\be
H_{\a\beta}=\frac{\kappa v_2 \delta_{\a\beta}-\gamma_{\a\beta}}{1+\kappa v_2},
\ee
where $\kappa$ is taken to be $1$, numerical values of $\gamma$ and $\delta$ are taken from \cite{Abelev:2009ad}, and values of $v_2$ are taken from \cite{Agakishiev:2011eq}. 
We see that by adjusting the $\tau_B$ parameter, the CME signal from our model is in a good agreement with that from experiments. And with the same $\tau_B(\simeq 1.65\text{fm})$, we predict the signal for Ru and Zr, which are larger than that of Au, due to the square of the scaling of $\mu_e(\t_f)$ as $Z^2 A^{-2(\zeta+\frac89)}$, with roughly $\zeta\simeq\frac14$, as we discussed in Section~\ref{scaling}.

\section{Conclusion}

We have calculated axial charge evolution using stochastic hydrodynamics model, and used it to get chiral magnetic effect in off-central collisions of AuAu, RuRu and ZrZr. By matching results from our model with background subtracted experimental data, we fix the relaxation time for magnetic field. We use the same relaxation time to make prediction for CME signal for collisions of RuRu and ZrZr. Two interesting results have been obtained in our analysis.

Firstly, while the axial charge and vector charge are coupled through chiral magnetic effect and chiral separation effect, we found the influence of vector charge to axial charge is negligible at top RHIC collision energy. This allows us to decouple the evolution of axial charge from the vector charge.

Secondly, we study the centrality and system size dependences of the CME signal. The initial chiral imbalance $\m_{A0}$ is found to have only weak centrality dependence. The centrality dependence of the CME signal comes mainly from the magnetic field and the QGP volume factor. As for the system size dependence, although larger system gives enhanced magnetic field and chiral imbalance, the electric charge asymmetry characterized by $e\m_e$ is suppressed due to larger volume factor. Consequently we found larger absolute charged particle correlation in isobar collisions than that in AuAu collisions.

The present study readily generalizes to collision of large nucleus at higher energies where we expect Bjorken flow approximation is still good. It would be interesting to see if the energy dependence matches with current experiment data at different energies. 
At lower energies, the Bjorken flow approximation becomes inaccurate. A possible approach is to implement the stochastic noises numerically in the existing AVFD model. We will report studies along this line in the future.

\section{Acknowledgments}
We are grateful to Huanzhong Huang, Guoliang Ma, Dirk Rischke, Gang Wang for discussions. G.R.L also acknowledge Institute for Theoretical Physics at Frankfurt University for warm hospitality where part of this work has been done.

\appendix

\bibliographystyle{CPC}
\bibliography{cmehic}

\end{document}